\documentstyle[epsfig,times,12pt]{article}
\def\be{\begin{equation}}
\def\ee{\end{equation}}
\def\br{\begin{eqnarray}}
\def\er{\end{eqnarray}}

\begin{document}
\title{A Tempt To Measure Reality}
\author{Bhag C. Chauhan{\footnote{e-mail: $<chauhan@cftp.ist.utl.pt>$  Homepage: $ http://cftp/~chauhan/$}}\\
\it Centro de F\'{i}sica Te\'{o}rica das Part\'{i}culas (CFTP)\\
Departmento de Fisica, Instituto Superior T\'{e}cnico\\
Av. Rovisco Pais, 1049-001, Lisboa-PORTUGAL}
\date{\today}
\maketitle

\begin{abstract}
Despite the extraordinary successes the two great bastions of $20^{th}$ century science (Quantum Theory and General Relativity) are troubled with serious conceptual and mathematical difficulties. As a result, further growth of fundamental science is at stake. 
Is this the end of science? Optimistic answer is ``NOT''! In this work, it is argued that science must continue its cruise, but with anew strategy -- a thorough recourse into the grass-root level working of science is required. 
In fact, all the scientific methods are based upon our sense perception, which keeps the outer physical universe as a separate entity, that is something quite independent of the observer. Basically, it is the observer -- the knower (human mind) -- which makes perception possible. It pretends a person or scientist to recognize or refute the existence of an object or a phenomenon.
It is also tempted to evince that working of human mind is epistemically scientific and can, in principle, be completely deciphered. It's inclusion in scientific theories, although tedious, can certainly spark a revolution in our understanding of nature and reality.
\end{abstract}

\newpage
\section{Introduction and Motivation}
Human having a curious mind has been continuously understanding the phenomenal world out there, and his perception of reality is gradually improving over the centuries. Science is a versatile tool which makes human to see the reality at a close-up. The great new truth revelations by the unbroken diligence of scientists for the centuries have dramatically transformed human's view about himself and his place in the universe. 
The spectacular success of science particularly in the last century has ascertained human to consider it to be a self-contained world-view independent or inclusive of its entire philosophical foundation.

Unfortunately, there are several difficulties in our theories which obstruct us to see the true picture of nature and reality. 
Along with the scientific knowledge, we have also gathered speculations, debates, and confusions. 
It seems as we are extending the radius of our scientific knowledge, so have been increasing the circumference of our ignorance and the truth is becoming more and more dispelled. 
After struggling for several years a desponded scientist, Albert Einstein, who have a number of pioneering contributions to the development of modern science, uttered: ``I used to think when I was young that sooner, or later all the mysteries of existence would be solved and I worked hard. But now I can say that the more we know, the more our existence turns out to be mysterious. The more we know, the less we know and the more we become aware of the vastness... Science has failed in de-mystifying existence, on the contrary it has mystified things even more.''

On the other hand, the protagonists have overlooked the fact, as for them the important thing is how science could be utilized to increase the physical ease of life. They are least concerned about the problems in the fundamental science, since they have their own targets. They even don't hesitate to apply the available scientific knowledge for the purpose of human destruction. They think that they can live more happily in this way. Ironically, they have led the whole human society onto the same vision. As a matter of fact, a soaring level of human happiness has been resulting to the endless lusty desires, which have given birth to a restless world with several problems. The green-earth-environment has been polluted, a massive development of nuclear weapons is taking place. The countries having not enough food-stuff to feed their hunger, but do sustain far-flying dreams to create missiles and atom-bombs. 

All the crisis that modern science and humanity are facing today, clearly show that there is something wrong in our conventional way of thinking, the way we understand nature and our relation with it. The thinking of protagonists, who are the directors of the minds of the whole social setup, is the product of the existing world-view; meaning that modern scientific knowledge is inadequate to educate them in the right direction. So there is a need to understand fully the working of the conventional method of scientific studies. In this work, the subtle problems of modern science in understanding nature and reality are exposed and a candidate solution has been proposed.

The sections are divided as follows: In section 2, advancements and difficulties of quantum theory and modern cosmology are discussed. A secret dream of scientists for centuries -- Theory of Everything (TOE) -- seems hard to be fulfilled is shown in section 3. A recourse into the foundation of science is attempted in section 4 supplemented with a short discussion in section 5. Finally, the conclusions are presented in section 6. 

\section {Human Quest For Reality}
The studies of archaeological survey show that human mind has been always agitating and susceptible to the external stimuli. 
The numerous developments, as found in the excavations and investigations, dating back to the time roughly 35,000 years ago, show the beginnings of the emergence of a reflective consciousness. 
Records of stone tools, burial sites, cave art, and of migration patterns evince that a first awakened human culture was born in these glimmerings of personal and shared awareness. 
There happened a dramatic change in the view of reality and human identity at about roughly 10,000 years ago when our ancestors shifted from a nomadic life to a more settled livings in villages and farms; and then followed by a rise of city-states and the beginnings of civilization at roughly about 5,000 years ago.
  
A more recent revolution in the human awakening is clearly visible through the vast existing literature, museums and the developments in all the pathways of our life. The birth of science happened roughly 300 years ago with a radical dynamism and materialism of the industrial era. 
Science is based most assuredly on analysis, that is, scrutinizing every phenomenon and examining every part of it and finding out how it came about. 
The scientific revolution gave a totally different awareness to the human understanding of reality -- all aspects of life have vividly changed with it, including the work that people do, the ways they live together, how they relate to one another, and how they see their role in society and place in the universe.

The humanity's prevailing paradigm is changed again by another radical world-view, which was kicked off in the beginning of $20^{th}$ century with the emergence of a new vision of matter and universe. The modern concept of matter in subatomic physics from {\it quantum theory} and the new concept of space-time from the {\it theory of relativity} are totally different from the one, of which we were traditionally used to. These new explorations have changed our conception of the universe as whole with life in it.

The evolution in the life style of human is an implication of his desire and necessity of physical comfort and to know more and more about nature and physical reality.

\subsection {Looking Deep Into The Matter}
Quantum theory grew out of a series of anomalies in the picture of matter and light offered by classical physics -- in particular associated with black-body radiation, the photo-electric effect, and the need to devise a model of the atom consistent with the newly discovered sub-atomic particles. Without quantum physics, we are unable to explain the behavior of solids, the structure and function of DNA, super-conductivity, properties of super-fluids, and burning of stars etc...
There is no doubt that quantum theory has been one of the most profound discoveries of the $20^{th}$ century development of science. Indeed, this theory has become dramatically successful in order to explain the experimental results, which were, otherwise, impossible to understand in the classical formalism.  
It is generally agreed that quantum theory is, if not a complete explanation, at least a great step forward in the measurement of reality.   

Despite the extraordinary successes this theory has been plagued by conceptual difficulties. The debate about the relation of quantum mechanics to the familiar physical world continues. It is not at all clear, what this theory is about and what does it, in fact, describe? \cite{sgold}.   
From its inception the theory had has a ``measurement problem'' with the troubling intrusion of the observer \cite{Wheeler} in experiments.  
An Irish physicist J.S. Bell\footnote{Who became well known as the originator of Bell's Theorem, regarded by some in the quantum physics community as one of the most important theorems of the $20^{th}$ century.} has quoted in his book, "Speakable and Unspeakable in Quantum Mechanics" \cite{bell}: ``... conventional formulations of quantum theory, and of quantum field theory in particular, are unprofessionally vague and ambiguous. Professional theoretical physicists ought to be able to do better.''

Albert Einstein was not at all comfortable with the foundation and working of quantum theory, despite the important role he had played in the development of this theory (he was awarded the Nobel Prize for discovering the photo-electric effect). 
Nevertheless, it is a general conviction among the scientists that Niels Bohr\footnote{A Danish physicist who received Nobel Prize in 1922 for his services in the investigation of the structure of atoms and of the radiation emanating from them.} (founder of Copenhagen interpretation of quantum mechanics) vanquished Einstein in their famous, decades-long, debate \cite{debate}. On the other hand, till the end of his life, Einstein continued to pretend that perhaps the quantum mechanical description is not the whole story.
Erwin Schrodinger, one of the founders of the quantum theory and who is also known as the father of wave function\footnote{A continuous function that contains all the measurable informations about the quantum particle.}, was one of the most acerbic critics of the theory. He ultimately found this theory as impossible to believe.

\vspace{0.4truecm}
There are several {\it mysteries, puzzles and paradoxes} in Quantum Mechanics:

\vspace{0.3truecm}
\begin{small}
\noindent
However, the Schrodinger equation is perfectly linear, propagates continuously in time, but collapses discontinuously when a particle interacts with a classical system at the event of measurement. In fact, there is no dynamical description for the {\it collapse} of the wave function.

\vspace{0.4truecm}

\noindent
A quantum system is described with a complex wave function ($\psi$) which is an {\it abstract} entity, but whose squared value ($|\psi|^2$) represents its physical properties. This gives a probability distribution for where discrete particles may be found once the wave function is collapsed by an act of observation.

\vspace{0.4truecm}

\noindent
Quantum particles can have spooky connections: According to theory, they can communicate over vast distances in an instant, which gave rise to the famous EPR paradox \cite{EPR} and Bell's theorem \cite{bell1}. This ghost action violates the principle of the limitation of the velocity of light in relativity theory and the principle of causality. 

\vspace{0.4truecm}

\noindent
There is a profound relationship between measurement and reality, where reality depends heavily on the measurement techniques. Observation would create a different kind of reality than what existed independently. In other words, reality existed in a different way while under observation than it did in itself.

\vspace{0.4truecm}

\noindent
According to the ``Principle of Superposition'' the Schroedinger's Cat inside a box \cite{scat} is neither {\it dead} nor {\it alive}, but a superposition of these two states. The wave function thus contains the superposition of all possible states of a system until it is observed. 

\vspace{0.4truecm}

\noindent
A quantum particle can behave as a {\it wave} as well as a {\it particle}; e.g., in photo-electric effect it shows its particle nature whereas in a double slit experiment it behaves like a wave.

\vspace{0.4truecm}

\noindent
It is fundamentally impossible to measure the key physical quantities, in certain pairs, e.g., position and momentum, simultaneously to any desired degree of accuracy. Attempts to increase the precision of one measurement, result in less precise measures of the other member of the pair: ``Principle of Uncertainty.''
\end{small}

\vspace{0.6truecm}

There were several attempts to falsify this theory on conceptual and experimental grounds, e.g., Albert Einstein with the collaboration of Boris Podolsky and Nathan Rosen proposed a gedanken experiment [EPR Experiment]\cite{EPR} as an attempt to show that quantum mechanics was somehow not complete and that the wave function does not provide a complete description of physical reality. However, they left open the question of whether or not such a description exists.
J.S. Bell proved mathematically through an inequality, famously known as ``Bell's Inequality''\cite{bell}, that quantum mechanics does violate special relativity by allowing instantaneous interactions across even the cosmological distances.  This weird fact has been observed in an experiment by A. Aspect {\it et al.} in 1982 \cite{aspect}. In another attempt, Erwin Schroedinger fabricated a thought experiments \cite{scat}: {\it Cat-in-a Box}, where the future of a cat paradoxically depends on the random decay of a radioactive atom. Astonishingly, according to quantum theory, the hapless cat is neither dead nor alive but in a state of superposition of the two possibilities, before to be seen actually -- which is ridiculous and hard to swallow. In this way, the theory which helps us to look deep into the matter is resting on the serious conceptual difficulties.

\subsection{Looking Deep Into The Cosmos} 
Einstein's theory of general relativity gave a new vision toward the understanding of the dynamics of heavenly bodies and the origin and evolution of universe.
In modern cosmology the most popular theory today we have is the {\it big-bang theory} \cite{bigb}. According to this theory there was nothing before the big-bang and all the space-time must have originated there and then (``t=0''). No matter/ energy could exist before this bang, as there was no space and time for it to be in. The theory further describes that this universe evolved from a dense, nearly featureless hot gas and that is expanding and cooling continuously. 

Scientific evidences strongly support that the universe had a definite beginning a finite amount of time ago and also prove that the early universe was very hot and that as it expands, the gas within it cools. There are three important observations strongly supporting  the big-bang model: 1) The expansion of the universe observed in 1929 by Edwin Hubble. 2) The abundance of the light elements H, He, Li (according to the theory these light elements should have been fused from protons and neutrons in the first few minutes after the big-bang). 3) The discovery of the Cosmic Microwave Background (CMB) radiation. The theory claims that the CMB radiation is the remnant heat leftover from the big-bang and the frequency spectrum of the CMB should have a blackbody radiation form. This was indeed measured with tremendous accuracy by an experiment on NASA's COBE satellite. The recent Wilkinson Microwave Anisotropy Probe (WMAP) mission reveals conditions as they existed in the early universe by measuring the properties of the CMB radiation over the full sky \cite{WMAP}.

Although, this theory has passed some scientific tests, there are still many more trials, which it must undergo successfully.
In the context of a recent test of this theory, John Bahcall -- a leading solar-neutrino physicist and astrophysicist -- writes \cite{r6}: ``I am happy that the big-bang theory passed this test, but it would have been more exciting if the theory had failed and we had to start looking for a new model of the evolution of universe''. 
In fact, there are many domains of modern cosmology which are far from being settled. The theory is silent about what banged, why it banged, or what happened before it banged. Despite its name, the big-bang theory does not describe the {\it bang} at all. The biggest problem of the big-bang theory of {\it the origin of the universe is philosophical} -- perhaps even theological -- what banged and why it banged!  

The philosophical base of the theory stands as embarrassing situation for the scientists. Robert Jastrow -- the first chairman of NASA's Lunar Exploration Committee -- himself admitted \cite{r5a}: ``Astronomers try not to be influenced by philosophical considerations. However, the idea of a universe that has both a beginning and an end is distasteful to the scientific mind''. 
To avoid this initial difficulty the idea of {\it singularity} was introduced in which the universe expands from a singular point and collapses back to the singular point and repeats the cycle indefinitely \cite{r5}. The idea was appreciated to avoid the philosophical, rather theological, base of the theory, but the available experimental evidences indicate that this type of oscillating universe is a physical impossibility. The facts and recent results suggest the geometry of the universe is flat and will expand forever \cite{r8,WMAP}. So, the attempts behind this idea to avoid philosophical or theistic beginning of the universe all fail \cite{r9}.

The philosophical origin of the big-bang is hard to quit even in the current attempts that are being made through a highly speculative theory of unification of quantum mechanics with gravity: ``Quantum Cosmology''. It must be noted that the meaning of ``t=0'' is highly contextualized by the assumptions and limitations of big-bang theory. In the alternative theories like quantum cosmology, they may well address the problems like ``t=0'' but the underlying philosophical ideas about space, time, matter and causality, far from being eradicated, might re-emerge in new and distinctive patterns and which will lead to further questions.

\section{TOE Project}
Science works under the principle of economy of understanding nature \cite{occamr}: when multiple explanations are available for a phenomenon, the simplest version must be preferred. The logical description of a vast range of physical phenomena from a few basic principles, rather than the memorization of a large number of isolated facts or formulae. Such economy is the strength of modern analytical science.

Scientists have a secret dream to expound nature in the {\it simplest version}. They want to explain all phenomena in the universe with the minimum number of particles interacting with a single interaction. Search for such a TOE is like the quest for the {\it Holy Grail} in the Middles Ages.

TOE is a beautiful contemplation of theoretical physics and mathematics that fully explains all the known and unknown -- everything in entire universe including life -- with a single unified equation.
Search for such a theory has started from the idea proposed by Isaac Newton. According to him, one great theory might exist that would link all the other known theories and this Grand Unified Theory (GUT) would be able to describe everything including life in the entire universe.

Science has traversed a long way since the time of Newton, and other physicists, including Albert Einstein, began to realize this beautiful idea of unification. This idea became more popular after the revolutionary work of James Clerk Maxwell (1831-1879): The first theoretical unification of the two physical phenomena -- electricity and magnetism -- into one all-encompassing framework. 
The next great step was the success of Quantum Electrodynamics (QED) theory (the integration of electromagnetism and quantum mechanics)\footnote{This landmark work in the direction of GUT earned Richard Feynman, Julian Schwinger, and Sin-itiro Tomonaga the Nobel Prize for physics in 1965.}.
On the same lines, the unification of electromagnetic and weak nuclear forces known as Electro-Weak theory (EW)\footnote{In 1979, Sheldon Glashow, Abdus Salam, and Stephen Weinberg were given the Nobel Prize for this work.} took place. 

In order to find the most promising road to a GUT there are continuous endeavors to unify all the forces of nature. Mathematically elegant Kaluza-Klein theory does indeed succeed unifying gravity and electromagnetism in a 5 - dimensional formalism. Many ideas of this theory are the basis for the several modern unified theories that can by themselves form a GUT, namely {\it string theory, super-gravity and loop quantum gravity}.
Scientists want to see this theory (GUT) as an unification of general relativity theory, that describes the large scale structure in the universe and quantum theory, that studies the microscopic structures.

Although, the idea of unification seemed quite rewarding, yet the several difficulties at theoretical, experimental and phenomenological level have faded away the hope of realization of this elegant dream: What once seemed very near on the horizon may be further off than imagined.
Much of the difficulty in merging these theories comes from the radically different assumptions that these theories make on how the universe works. On the one hand, in conventional GUTs like SU(5) physical particles exist in the flat space-time of special relativity, whereas on the other hand in general relativity space-time is curved and that changes by the motion of mass. 

Noticing that a class of GUT quantum theories proposed in 1980's and later \cite{gut} couldn't pass even the first test in the laboratory: In 1999, Superkamiokande experiments reported that they had not detected proton decay as predicted by the GUTs \cite{sk99}. Also none of the generic predictions of these theories, the existence of topological defects such as monopoles, cosmic strings, domain walls etc... has been observed yet. As a result, not a single such quantum theory is currently universally accepted. 

On the other hand, the very complexity of Einstein's general relativity was first noted by himself as leading to a very serious impediment on its further development. In fact, after publishing his famous paper in 1916, he conceded that this arose from the mathematical difficulties involved in the complexity of its nonlinear coupled equations and their huge number of terms.                          In 1952 he expounded it as an acute frustration: ``The generalization of the theory of gravitation has occupied me unceasingly since 1916.''

Obviously, at this stage of debates and confusions, unification of all the four interactions is extremely difficult.
   
\section{\it A Recourse and A Proposal} 
With the help of science we observe, describe and establish the truth on ocular demonstration and verify it with experiments which anyone may undertake without the least faith in ultimate results.
If we believe that science must cruise in a consistent way towards reality and not get lost in the so-called confluence of chaos and confusions, then  
all the above discussed facts and difficulties faced by modern science and humanity put a big question-mark (?) on the grass-root level working of science. 

A serious flaw is resting in the foundation of physics \cite{SIT}. And there is a serious need to understand completely the working of the conventional method of scientific studies.
Recalling that all scientific researches are based on {\it Cartesian Partition} approach: Relying upon ordinary sense perception, which keeps the outer physical universe as a separate entity, to be an independent existence, that is something quite independent of the observer. Notice that here we are separating the real observer from the observation and only relying upon the sense perception of human body.

In order to make a perception possible there must be a subject --the knower-- who can observe a phenomenon or an event with the help of a connecting principle.
In fact, it is not the physical part of human brain which acts as the observer (the knower) and makes the perception possible, but there exists a subtle playback entity; a consciousness being -- {\it Mind}. The human mind is the doer, the observer which interprets the messages collected from outside by the brain with the help of sense organs and instruments. Human mind is a part of nature and an essential component of our observations, and there is no point in eliminating it from the measurement process.

\subsection{\it Some Favorable Conjectures} 
Let's ponder over some noted issues which are revolving around or converging into the entity: human mind. They show human mind as an inevitable target of scientific contemplation in order for the further growth of fundamental science.
  
{\bf 1.} Recalling that the classical physics, the study of macroscopic world, is based on the principle of {\it Cartesian Partition}. Given that in the last century's development of science, there came up a well corroborated fact that the classical physics is an incomplete understanding of nature. This suggests that our scientific research based upon 'Cartesian Partition' approach must be incomplete or erroneous.  

Nevertheless, to a first approximation this approach is fine. It is simple and workable, as evidenced by the success of science since its birth. However, as we enter deep into the matter -- with quantum theory, still clinging to this approach -- then we have to have always face the weird responses of mother nature to our questions. Some people called this as the intrusion of observer in the act of measurement \cite{Wheeler}. On their lines, it is straightforward to argue that this acute ascendancy of observer in the measurement process experimentally confirms the prevailing importance of the functioning of human mind -- the real observer -- in the definition of reality. In this way it's scientific contemplation is inevitable. 
Interetingly, the relationship of mind and matter, which eliminate Cartesian partition, has already been widely explored in literature \cite{mind_matter}. 

{\bf 2.} It has been witnessed by the gradual growth of science, the several interfaces among its different branches have been emerging out. For example, genetic engineering and the associated reproductive technologies on plants, animals and human have brought forth ethical issues calling social scientists and environmentalists for greater regulation to hold. New disciplines like Bio-Physics, Bio-Chemistry, Ecology, Astro-Particle Physics etc... are already in their establishments. It appears that finally all the branches of science, including social and behavioral sciences, are going to meet at some point of time. There can no longer be ``pure'' science -- every branch of science reacts with others \cite{CERN_C}. 

As all the fields of science are developing, they are converging, and the mysteries of human mind in almost each of them are coming up into light \cite{capra}. This clearly implies a global necessity for understanding the functioning of human mind. It seems obvious that a scientific contemplation of human mind could bring all the fields of knowledge on a common platform, and certainly impel human awareness a leap forward. 

{\bf 3.} As has been already mentioned above, modern scientific advancement has influenced all the sectors of our day-to-day life including our thoughts and culture. There is no doubt that along with the enormous physical comforts, mental restless and all the problems at personal, social, and global levels that we face today are also related to our scientific understanding of nature.
So, there exists essentially a crisis of understanding of our own minds and nature -- a crisis of true perception of reality. In order to maintain both peace and prosperity together we have to understand the functioning of our minds and learn to eliminate the causes which promote human toward destruction.

According to Melvin Calvin \cite{CERN_C}, a Nobel laureate in chemistry, that it is apparent that for the welfare of mankind, scientists must understand the basic knowledge of other fields than their own, and, in addition, must understand world about them in terms of the humanist as well. And, conversely, the student of humanities must understand the interrelationships of his own specialty (for example, of urban planning, with the humanitarian, or aesthetic, provisions for peace of mind and of environment) as well as the relationship of his specialty to new knowledge advanced in the area of science.

{\bf 4.} We know that our thoughts and emotions do influence our brain chemistry and other biological activities, yet for no significant reasons we don't treat them in the definition of reality. 
The mind composed of thoughts and emotions do influence our observations and measurements. In this way mind-independent measurement of reality is erroneous and incomplete. So the functioning of mind must be incorporated in our scientific methods in order to understand the true picture of reality.   

{\bf 5.} The most creative physicists have always emphasized that human consciousness (mind) is at the foundation of the scientific method behind physics.
According to American physicist Eugene Wigner: ``The next revolution in physics will occur when the properties of mind will be included in the equations of quantum theory''. Luis De Broglie -- who proposed the idea of the wave-nature of particle -- said: ``The structure of the material universe has something in common with the laws that govern the working of the human mind''. Erwin Schroedinger felt deeply that human mind is a sole constructor of all the observations and quoted as: ``Our picture of the world is, and always will be, a construct of the mind''.  In order to construct reality mind has been thought responsible for the collapse of wave function \cite{collapse}. The idea give rise in some cases to a defense of freedom of will.   

{\bf 6.} A radical change in the human understanding of nature and objective reality is expected by unveiling the mysteries of human mind. This revolution of human consciousness is quite probable and supported by the several leaps already happened in the history of evolution \cite{SIT}.

In summary, there are some facts which are interpreted as the compelling evidence for the scientific contemplation of human mind. And there is a need to expand our world-view in order to include human mind in the definition of reality.

\section{\it Discussion}
Science is an objective study of matter and its interactions based on a tacit assumption that information about a physical system can be acquired without influencing the the system's state. However, the ``information'' is regarded as unphysical, a mere record of the tangible, material universe essentially decoupled from the domain governed by the physical laws.  
 
The Cartesian approach of science finds it extremely difficult to include an entity, in the theory, which is just a subjective experience. 
So, in order to contemplate human mind scientifically, first of all we need to check whether it (mind) is objective in nature or not.
If human mind is purely subjective in nature, as per general prejudice, then it would certainly be impossible to contemplate it scientifically. However, on the other hand, the objectivity property of mind can dramatically simplify the understanding of it and thereby facilitate its inclusion in scientific theories. 

Metaphorically, human mind can be understood as a lake of water. The surface of lake is the dividing line between sleeping and waking state of consciousness. Below the surface is the subconscious and unconscious mind, and on the surface is the conscious state of mind.
It is argued here that human mind is subjective only in its superficial layers which retain and defend the individuality of a person. However, the objectivity is there in the whole remaining part of it, as because all humans are alike in the deep down, having the similar qualities and traits: quest for truth and happiness, an urge to be alive, and seek justice, sense of guilt, love and compassion, kindness etc... and all those abilities necessary to carry life. For example, if you ask a terrorist: why he was shedding blood of innocent lives. His answer is simple and straightforward: ``I don't like to do but am forced by the circumstances.'' This clearly shows the objectivity of mind below the superficial {\it terrorist-mask}. 

Traditionally, using statistical analysis, the three factors genetic, environmental, and the product of both components have been recognized as responsible for all the behavioral variations.
It has been accepted since a long time by the biologists that the genes, the environment, and the interaction between them orchestrate the human behavior \cite{gene}. 
Which shows that only these three factors contribute to shape the personality of a child and offer a shallow individuality to him. So, in this way, the subjectivity of human mind at the superficial levels (although not well-understood yet) can be scientifically contemplated. It is up to the further advancements of neuro-studies, genetics, and in behavioral sciences to realize this dream. 

For further elaboration of this concept, let's perform a thought experiment:  

\subsubsection*{Mind: A Compact Disc}
The mind of every new-born baby is a {\it formatted and blank} compact disc ({\it Mind-Disc}) produced by the same company (nature) through a certain franchised firm (parents). To call it {\it formatted}, means that there already exist some genetic and biological instructions like, a certain mental level and some specific traits, e.g., cry for food, sucking of nipple, excretion, breathing, pumping of heart and all of those are necessary to carry life. The Mind-Disc is called {\it blank} because it is like a clean slate, whatsoever you want to print on it, you can. 

So, let's take the mind of a new-born baby as a fresh sample to test. As child grows up social factors start shaping his life. It is a noted fact that the parents, teachers and society are the dominant factors responsible for an innocent child to become a criminal or a gentleman in his later life. Evidently, this example, to a far extent, supports the objectivity of human mind and thereby establishes that determinism is working on the most part of it. The other factors influencing the human behavior are genetic and biological (as a format of the mind-disc set off by nature). If we could contemplate them scientifically, a total objectivity of human mind can be established.

\section{Conclusions}
Although, classical physics has failed to explain the dynamics of the microscopic particles, yet, ironically, modern scientific researches are based upon the prejudice posed by classical physics -- Cartesian Partition -- keeping the outer physical universe as a separate entity. Indeed, the quantum physics experiments have knocked the door of a new paradigm through the troublesome intrusion of observer -- human mind -- in the act of observation. The prevailing role of observer, as suggested by the quantum physics experiments, is one of the physical proofs substantiating the idea proposed in this work as a solution of the problem in hand. 

Several observed facts have been analyzed and conjectures have been presented, which favor the scientific contemplation of human mind in order to facilitate the further growth of science and humanity. The objectivity property of human mind has been brought into light. As a result of which, it would be dramatically simpler to comprehend it in the scientific theories.  
Although, the scientific contemplation of human mind is a great challenge for the scientists yet, it there are strong possibilities that a fruitful collaboration of the experts from all disciplines of life could spark and may facilitate the accomplishment of the holistic cause.   

In the light of the indicative conjectures, once some concrete steps are made in this direction, the solutions to the various problems related to the difficulties and growth of modern science, and peace and prosperity of humanity would start showing up.
Finally a paradigm-shift is evinced, which has the potential to dramatically transform our view of reality, identity, social relationships, and human purpose.

\section*{\it{Acknowledgments}}
The work was supported by Funda\c{c}\~{a}o para a Ci\^{e}ncia e a Tecnologia through the grant SFRH/BPD/5719/2001.



\begin{thebibliography}{99}
\bibitem{sgold} S. Goldstein, ``Quantum Theory without Observers -- Part 1-2'', Physics Today, p.42, March 1998 and p.38, April 1998.
\bibitem{Wheeler} J. A. Wheeler and W. H. Zurek, ``Quantum Theory and Measurement'', (eds., Princeton U. P., Princeton, 1983).
\bibitem{bell} J. S. Bell, ``Speakable and Unspeakable in Quantum Mechanics'', (Cambridge University Press, 1987). 
\bibitem{debate} Bohr v Einstein: The debate between Bohr and Einstein over the interpretation of quantum theory began in 1927 at the fifth Solvay Conference of physicists and ended at Einstein's death in 1955. 
\bibitem{EPR} A. Einstein, B. Podolsky and N. Rosen, ``Can quantum-mechanical description of physical reality be considered complete?'',  Physical Review {\bf 47}, 770 (1935).
\bibitem{bell1} J. S. Bell, ``On the Einstein Podolsky Rosen paradox'' Physics {\bf 1}, n3, 195 (1964). 
\bibitem{aspect} A. Aspect, Dalibard, G. Roger, ``Experimental test of Bell's inequalities using time-varying analyzers'' Phys. Rev. Lett. {\bf 49} 1804 (1982).
\bibitem{scat} E. Schroedinger, ``Quantum Theory and Measurement'' Naturwiss. {\bf 23}, 807 (1935), translated to English ed. J.A. Wheeler and W.H. Zurek, (Princeton Univ Press, 1983).
\bibitem{bigb} E. Kolb and M. Turner, ``The Early Universe'' (Westview Press, 1990).
\bibitem{WMAP} C. L. Bennett {\em et al}., ``First year Wilkinson Microwave Anisotropy Probe (WMAP) observations: Preliminary maps and basic results'' Astrophys. J. Suppl. {\bf 148} 1 (2003).
\bibitem{r6} J. Bahcall, ``The Big Bang is Bang On'' Nature {\bf 408} (2000).
\bibitem{r5a} R. Jastrow, ``Until the Sun Dies'' (New York: W.W. Norton, p.31, 1977).
\bibitem{r5} J. Gribbin, ``Oscillating Universe Bounces Back'' Nature {\bf 259} (1976).
\bibitem{r8} J. Silk, ``The Big Bang'', (San Francisco, CA: W.H. Freeman, p.309, 1980); J. Gribbin, ``Genesis: The Origins of Man and the Universe'', (New York: Delacorte, p.316, 1981); R. Jastrow, ``God and the Astronomers'', (New York: W. Norton, p.123, 1978); E. Chaisson, ``Early Results from the Hubble Space Telescope'', Scientific American {\bf 266}, 6 (1992).
\bibitem{r9} H. Ross, ``The Fingerprint of God'', (Orange, CA: Promise Publishing, p.105, 1991).
\bibitem{occamr} Occam's Razor: A principle attributed to the 14th century English logician and Franciscan friar, William of Occam that forms the basis of methodological reductionism, also called the principle of parsimony. This has become a basic perspective for those who follow the scientific method.
\bibitem{gut} H. Georgi and S. Glashow, ``Unity of All Elementary-Particle Forces'', Phys. Rev. Lett. {\bf 32}, 438 (1974); Graham Ross, ``Grand Unified Theories'', (Benjamin-Cummings Publishing Company, 1985). 
\bibitem{sk99} Superkamiokande Collaboration (Y. Hayato et al.) Phys. Rev. Lett. {\bf 83}, 1529 (1999).
\bibitem{SIT} Bhag C. Chauhan, ``Science In Trauma'', archive: physics/0210088.
\bibitem{CERN_C} Melvin Calvin's ``The Impurity of Science'' referenced in CERN Courier {\bf 45}, 11 (2005).
\bibitem{capra} F. Capra, ``The Turning Point'' (Simon \& Schuster, New York, 1982); ``Uncommon Wisdom'' (London: FLAMINGO, 1989). 
\bibitem{collapse} A. H. Compton, ``The Freedom of Man'', (Yale University Press, New Haven 1935); ``Reinventing the Philosophy of Nature'', (Review of Metaphysics, p. 3, 1981); J. von Neumann, ``Mathematical Foundation of Quantum Mechanics'', (Princeton University Press, 1955); E. P. Wigner, ``The Scientist Speculates'', (I. J. Good, ed., Heineman, London, p.284, 1961); L. Bass, ``A Quantum Mechanical Mind-Body Interaction'', Foundation of Physics {\bf 5(1)}, 159 (1975); H. P. Stapp, ``Quantum Properties and Brain-Mind Connection'', Foundation of Physics {\bf 21(12)}, 1451 (1991); Mind, Matter and Quantum Mechanics, (Springer-Verlag, New-York, 1993); R. Penrose, ``Shadows of the Mind'', (Oxford University Press, Oxford, 1994).  
\bibitem{gene} Gene E. Robinson, ``Genomics: Beyond Nature and Nurture'', Science {\bf 304}, issue 5669, 397 (2004). 
\bibitem{mind_matter} D. Bohm, ``A New Theory of Relationship of Mind and Matter,'' Philosophical Psychology, {\bf 3}, No.2, 271 (1990); D. Bohm and B. J. Hiley, ``The Undivided Universe'', ch. 15, (Routledge, London, 1993); B. J. Hiley, ``Non-commutative geometry, the Bohm interpretation and the mind-matter relationship'', Computing Anticipatory Systems - CASYS 2000, ed. by D. Dubois, Springer, Berlin, p. 77, 2003); 
C. G. Jung and W. Pauli, ``The Interpretation of Nature and the Psyche'', Pantheon, New York, 1955); 
H. Atmanspacher and H. Primas, ``The hidden side of Wolfgang Pauli'' Journal of Consciousness Studies {\bf 3}, 112 (1996). 
\end{thebibliography}
\end{document}